# The transition from the ballistic to the diffusive regime in a turbid medium

Andre Yaroshevsky, Ziv Glasser, Er'el Granot and Shmuel Sternklar

Department of Electrical and Electronics Engineering, Ariel University Center of Samaria, Ariel, Israel
*Corresponding author: andrey@ariel.ac.il*

By varying the absorption coefficient and width of an intralipid- India ink solution in a quasi one-dimensional experiment, the transition between the ballistic and the diffusive regimes is investigated. The medium's attenuation coefficient changes abruptly between two different values within a single mean-free-path. This problem is analyzed both experimentally and theoretically, and it is demonstrated that the transition location depends on the scattering coefficient as well as on the measuring solid angle.

Photon transport in the transition regime between ballistic and diffusive propagation is the subject of much attention [1-4]. This regime holds valuable information on the nature of scattering.

In general, multiple scattering is an extremely complicated process. However, the specific route from ballistic to diffusive transport carries information regarding the most relevant approximation that should be used in the analysis, such as the single scatter approximation, coherent methods, Beer's law, the transport equation or diffusion approximations. One can also infer from the transition behavior the light properties which are dominant in the process.

Most of the published work indicates that the width of the transition regime is approximately several mean free paths (MFP) [2-6]. The term "MFP" is somewhat confusing. Usually it refers to the mean distance between successive collisions, which is the reciprocal of the scattering coefficient $\mu_s$. In the ballistic regime this is the most important parameter since any photon that experiences scattering is no longer ballistic. In the diffusive regime, where multiple scattering is the dominant process, the distance between successive scattering events is less important, whereas the significance of the scattering angle increases. In this regime a new length scale becomes dominant – the transport mean free path, which is the reciprocal of the effective scattering coefficient $\mu_s'$. This length scale can be understood to be the length in which the photon "forgets" its initial orientation. If $\theta$ is the scattering angle in a single scattering event, and its mean cosine is $g \equiv \langle \cos\theta \rangle$ then $\mu_s' = (1-g)\mu_s$ [7], i.e., $(1-g)^{-1}$ evaluates the mean number of scattering events which leads to a loss of the initial orientation of the photons.

To investigate the transition from the ballistic to the diffusion regime, one can measure the transmitted light within a relatively narrow solid angle. As this angle decreases it becomes easier to distinguish between ballistic and diffusive photons. While the ballistic photons retain their initial direction, the diffusive ones, almost by definition, are scattered in all directions. Hence, in the diffusion approximation, the amount of diffusive photons which reach the detector is proportional to the measuring solid angle. On the other hand, if the angle is too narrow then only a negligible amount of diffusive photons will reach the detector.

Refs. 5 and 6 studied the ballistic-diffusive transition. In these works, the setups were arranged in a confocal geometry to prefer ballistic photons over diffusive ones. In our work we measure the photons directly without confocal configuration and the distinction is made by a distant aperture, whose diameter determines the measuring solid angle. Moreover, in this experiment the attenuation coefficient of the diffusive regime can be controlled by varying the absorption of the medium.

In the ballistic regime, the decay of the ballistic photon density $\rho_B$ is governed by $d\rho_B/dz = -(\mu_s + \mu_a)\rho_B$, where $\mu_a$ is the absorption coefficient. The diffusive photons are governed by the diffusion equation. Hence, their density in the stationary state can be described by (see, for example, Ref.7) $D\nabla^2\rho_D = c\mu_a\rho_D$, where $D$ is the diffusion coefficient, which can be written in terms of the scattering coefficients as $D = c/3(\mu_s'+\mu_a)$. Due to the three-dimensional nature of the geometry, a point source leads to a singularity in its' proximity, which will override the ballistic photons for short distances. To reveal the ballistic component, the source must be sufficiently wide so that the propagation will be quasi-one-dimensional. In this case the two transverse dimensions are degenerate and the diffusion equation can be approximated by

$$\partial^2 \rho_D / \partial z^2 = 3(\mu_s'+\mu_a)\mu_a\rho_D \qquad (1)$$

In cases where the solid angle $\delta\Omega$ of the measurement device is very small, then these two kinds of photons will be measured: the ballistic and the homogeneous diffusive ones. As a consequence the measured photons density is a sum of the ballistic photons and a portion of the diffusive photons, leading to: $\rho = \rho_B + (\delta\Omega/4\pi)\rho_D$, i.e.,

$$I/I_0 = \rho/\rho_0 = \exp(-\mu_s z) + (\delta\Omega/4\pi)\exp(-\mu_{eff} z) \quad (2)$$

where $\mu_{eff} \equiv \sqrt{3\mu_a(\mu_s'+\mu_a)}$, $I_0 \equiv I(z=0)$ and $\rho_0 \equiv \rho(z=0)$. Note that in diffusive media $\mu_a << \mu_s$ (otherwise it would have been an absorptive media), and therefore $\mu_a$ is neglected in the ballistic (left) term of Eq.2; however, since $\mu_s' < \mu_s$, it should not always be neglected in the diffusive (right) term (this is the case in our experiments). Eq.2 does not include a term for the quasi-ballistic photons, but as can be seen from the comparison to the experimental results their influence is negligible in our setup.

It should also be stressed that despite the fact that the diffusive part of Eq.2 was derived for an infinite medium it is an excellent approximation. A derivation for a finite slab would change the result by a negligible amount (less than $10^{-4}\%$). We therefore find a transition depth

$$z_c \equiv (\mu_{eff} - \mu_s)^{-1} \ln(\delta\Omega/4\pi) \quad (3)$$

For $z << z_c$ the radiation reaching the detector is mainly ballistic, and its' dependence on the sample's width is governed by the relation $I/I_0 \sim \exp(-\mu_s z)$.

Clearly, when $z < z_c$ the diffuse photons are not scattered isotropically [8], however, in this regime, as can be seen from our experimental results in the following section, their influence is negligible and therefore the isotropic assumption of Eq.2 is an adequate approximation.

Beyond the transition width, i.e., $z >> z_c$, the ballistic photons have a negligible contribution to the detected signal. Most photons have diffusive properties and therefore are governed by $I/I_0 \sim (\delta\Omega/4\pi)\exp(-\mu_{eff} z)$.

We believe that since $\mu_s >> \mu_{eff}$ the transition width $z_c$ depends mainly on only two parameters: the scattering coefficient $\mu_s$ (or, alternatively its reciprocal – the mean free path) and the measuring angle $\Omega$. That is, beside $\mu_s$ the transition width is a property of the experimental conditions, and not a property of the medium. Most importantly: it has no dependence on the transport mean free path (i.e., it is independent of the anisotropy parameter $g$).

In principle we can extend the width in which ballistic photons can be measured by decreasing the angle $\delta\Omega$, however, there is no point in decreasing the angle below the ballistic scattering angle, which is the Gaussian beam' diffraction angle

$\delta\Omega = \pi \operatorname{atan}^2(2\lambda/\pi d) \cong \pi^{-1}(2\lambda/d)^2$ in which case the transition depth cannot exceed

$$z_c^{max} \equiv \frac{2}{\mu_{eff} - \mu_s} \ln\left(\frac{\lambda}{d\pi}\right) \cong \frac{2}{\mu_s} \ln\left(\frac{d\pi}{\lambda}\right) \quad (4)$$

(since usually $\mu_s >> \mu_{eff}$). The ratio between the maximum transition depth and the mean free path is $2\ln(d\pi/\lambda)$.

Since in most optical experiment the beam's width is several millimeters and the wavelength is a few microns, then $2\ln(d\pi/\lambda) \sim 10$, which is consistent with the literature [1,3,4,6]. Note from (4) it is evident that we cannot "see" the ballistic image of a turbid medium whose attenuation exceeds $(\lambda/d\pi)^2$.

Another interesting feature of these finding is that in order to see a ballistic image of a medium the ratio between the medium's traversal ($d$) and longitudinal ($z$) dimensions must satisfy:

$\frac{d}{z} > \frac{d\mu_s}{2\ln(d\pi/\lambda)} = \frac{\lambda\mu_s(d\pi/\lambda)}{2\pi\ln(d\pi/\lambda)} > e\frac{\mu_s}{k}$, where $k \equiv 2\pi/\lambda$ and $e = 2.72...$. Since at $z = z_c$ the two terms $\exp(-\mu_s z_c)$ and $(\delta\Omega/4\pi)\exp(-\mu_{eff} z_c)$ are equal, it follows that at the transition point each term deviates in the logarithm scale from the linear curve by only a factor of 2. Therefore, at the transition point the intensity is

$$I/I_0 = 2(\delta\Omega/4\pi)^{\left(\frac{\mu_s}{\mu_{eff}-\mu_s}\right)} \cong \delta\Omega/2\pi \quad (5)$$

As $\delta\Omega$ increases the transition depth (3) decreases and the intensity at the transition increases correspondingly. The transition regime (determined by the length where the factor 2 is reduced to $\sqrt{2}$) is therefore $|z-z_c| < \Delta$, where $\Delta = |\mu_{eff} - \mu_s|^{-1} \ln(\sqrt{2}-1) \cong 0.88\mu_s^{-1}$ is half the transition width. This result (5) also explains why the measuring solid angle has to be very small so that $-\ln(\delta\Omega/4\pi) >> 1$, otherwise, the transition will be washed out.

A schematic presentation of the experiment is illustrated in Fig.1. Diluted intralipid, commonly used as a tissue phantom, served as the turbid medium. In all the measurements we used the same light source (840nm cw beam), the same intralipid concentration, i.e., the scattering coefficient was $\mu_s \cong 140 cm^{-1}$, and the transport mean free path was $\mu_s' \cong 28 cm^{-1}$. It should be emphasized that in the literature [9,10] the solution is considerably diluted, and therefore the transition effect appears only in very large samples (an exception are Refs.5 and 6 where the transition appears at short distance but for a solution of microspheres). To overcome this problem we had to increase the concentration considerably.

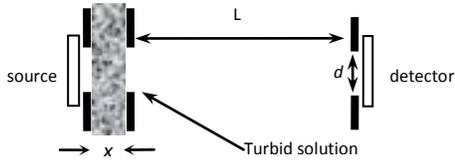

Figure 1: Schematic illustration of the experiment

We carried out four sets of experiments, each of which had a different concentration of ink, so that the absorption coefficient differed accordingly.

In each experiment the width of the solution was changed gradually and the transmission was measured through a distant aperture. The data was fitted to the equation

$$I/I_0 = \rho/\rho_0 = \exp(-\mu_s z) + \varepsilon \exp(-\mu_{eff} z) \quad (6)$$

In our experiment, the distance between the sample and the detector was $L = 30 cm$ and the diameter of the aperture in front of the detector was $d = 0.15 cm$, so that the theoretical solid angle was $\frac{\delta\Omega}{4\pi} \cong \left(\frac{d}{4L}\right)^2 \cong 1.5 \times 10^{-6}$, which is on the same order of magnitude as the experimental fit $\varepsilon \cong 4 \times 10^{-7}$. We cannot pinpoint the source of the factor 4 discrepancy between the theoretical prediction and the experimental result. We believe that it is probably due to the 1D approximation (Eq.5), which holds only approximately. A decay term such as $\exp(-\mu_{eff} z)$ holds only for an infinite 1D sample. For non quasi-1D samples, i.e., when the medium's thickness is not considerably smaller than the beams' traversal dimensions, this term is multiplied by a power law factor[7], which can be neglected only when the exponential attenuation is the dominant one, i.e. when the absorption is large.

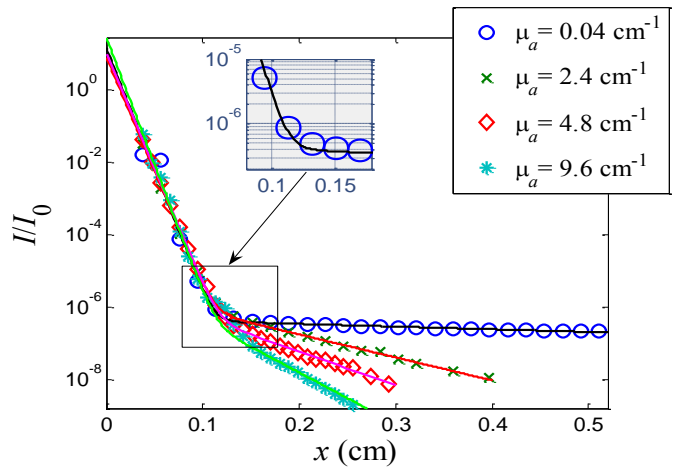

Figure 2: (color online) Transmission through diffusive medium vs. sample width, for various absorption coefficients. The transition area is shown in the inset.

In Fig. 2 we present the four experimental results and theoretical predictions on the same graph. For the three lower curves $\varepsilon \cong 4 \times 10^{-7}$. For the fourth case ($\mu = 0.04 cm^{-1}$), $\varepsilon \cong 3 \times 10^{-8}$. We believe that this deviation is due to the weak absorption coefficient, so that the quasi 1D approximation, i.e., Eq.5, fails. The two exponential regimes $z \ll z_c$ and $z \gg z_c$ are clearly seen, with an excellent fit to the theoretical prediction (solid lines).

To summarize, the transition from the ballistic to the diffusive regime was investigated in highly turbid tissue-phantom media. It was demonstrated that the transition between the two regimes occurs abruptly, where the total attenuation coefficients, which takes account of absorption as well as scattering, changes between two different values in a single mean-free-path. It is demonstrated that the ability to see the ballistic photons *is dependent on the solid-angle of the measurement device*, and is bounded by the ballistic beam's geometry.